\documentclass[prl,aps,twocolumn,superscriptaddress,nofootinbib,nopacs]{revtex4}
\usepackage{graphicx}  
\usepackage{dcolumn}  
\usepackage{bm}           
\usepackage{amsmath}
\usepackage{epsfig}
\usepackage{indentfirst}
\usepackage{psfrag}
\usepackage{subfigure}
\usepackage{amssymb}
\usepackage{tabularx,multirow}
\usepackage{color}
\usepackage[colorlinks,linkcolor=blue,citecolor=blue,urlcolor=blue,hyperindex,driverfallback=dvipdfm]{hyperref}
\usepackage[T1]{fontenc}
\usepackage{wasysym}

\def\rb{{\bf r}}      
      
    \def\kparb{{\bf k}_\parallel}
    
\def \Fot {F_{1\rightarrow2}} \def \Fto {F_{2\rightarrow1}} 
\def \Pot {P_{1\rightarrow2}}  
\def \Foo {F_{1\rightarrow1}}  
    
\begin{document}
\title{Time-modulated near-field radiative heat transfer}
\author{Renwen~Yu}
\affiliation{Department of Electrical Engineering, Ginzton Laboratory, Stanford University, Stanford, CA 94305, USA}
\author{Shanhui~Fan}
\email{shanhui@stanford.edu}
\affiliation{Department of Electrical Engineering, Ginzton Laboratory, Stanford University, Stanford, CA 94305, USA}
\begin{abstract}
We explore near-field radiative heat transfer between two bodies under time modulation by developing a rigorous fluctuational electrodynamics formalism. We demonstrate that time modulation can results in the enhancement, suppression, elimination, or reversal of radiative heat flow between the two bodies, and can be used to create a radiative thermal diode with infinite contrast ratio, as well as a near-field radiative heat engine that pumps heat from the cold to the hot bodies. The formalism reveals a fundamental symmetry relation in the radiative heat transfer coefficients that underlies these effects. Our results indicate the significant capabilities of time modulation for managing nanoscale heat flow.  
\end{abstract}
\maketitle

Electromagnetic fluctuations at nanoscale manifest in many important physics phenomena ranging from enhanced radiative heat transfer in near field \cite{VP07,SGS15} to equilibrium and nonequilibrium Casimir effects \cite{KMM09,MCP09}. In particular, near-field radiative heat transfer has been extensively studied \cite{PV1971,CG99,JMM05,KMP05,NSC08,FMV08,BB14,OLF10,KHZ12,RSJ09,KSF15,paper198,ZF16,ZGZ17,SLL15,PZB19,SZF16,BEK17,SOT14,BBJ11,AK22,paper286}, both for its fundamental physics implications, such as in vacuum friction \cite{paper157}, and for its application in energy technology, such as in thermophotovoltaic generation of electricity \cite{LCG06,BZF09,MLZ21}. Recent developments in electromagnetics indicate that significant new physics \cite{YF09,YSF16,YLX18,XMA14,LSS18} can arise when the permittivity is modulated as a function of time \cite{SSF16}. Motivated by these developments, there are emerging interests in exploring implications of time modulation in electromagnetic fluctuational phenomena \cite{BLF20,SRJ21,VL23,YF23,KPJ15,KLR15}. However, near-field radiative heat transfer between two bodies with time-modulations has not been previously explored.

In this work, we develop a rigorous fluctuational electrodynamics formalism to study
near-field radiative heat transfer between two bodies with time modulation. Our study shows that, compared with a corresponding system without time modulation, the near-field heat transfer at a given temperature bias can be significantly enhanced, suppressed, or even completely eliminated by time modulation. The complete elimination of near-field heat transfer, moreover, gives rise to the possibility of achieving a thermal diode with \textit{infinite} contrast between forward and backward biases. We also show that the direction of the heat flow can be reversed so that the system operates as an active cooler that pumps heat from the low to the high temperature bodies. Compared with the far-field case as studied in Ref.\ \cite{BLF20}, operating in the near field can result in the enhancement of cooling power density by five orders of magnitude. Our formalism also reveals a symmetry relation that underlies many of these novel effects as mentioned above.  The results point to significant opportunities in the explorations of time modulation in near-field heat transfer.

\begin{figure}
	\noindent \begin{centering}
		\includegraphics[width=0.48\textwidth]{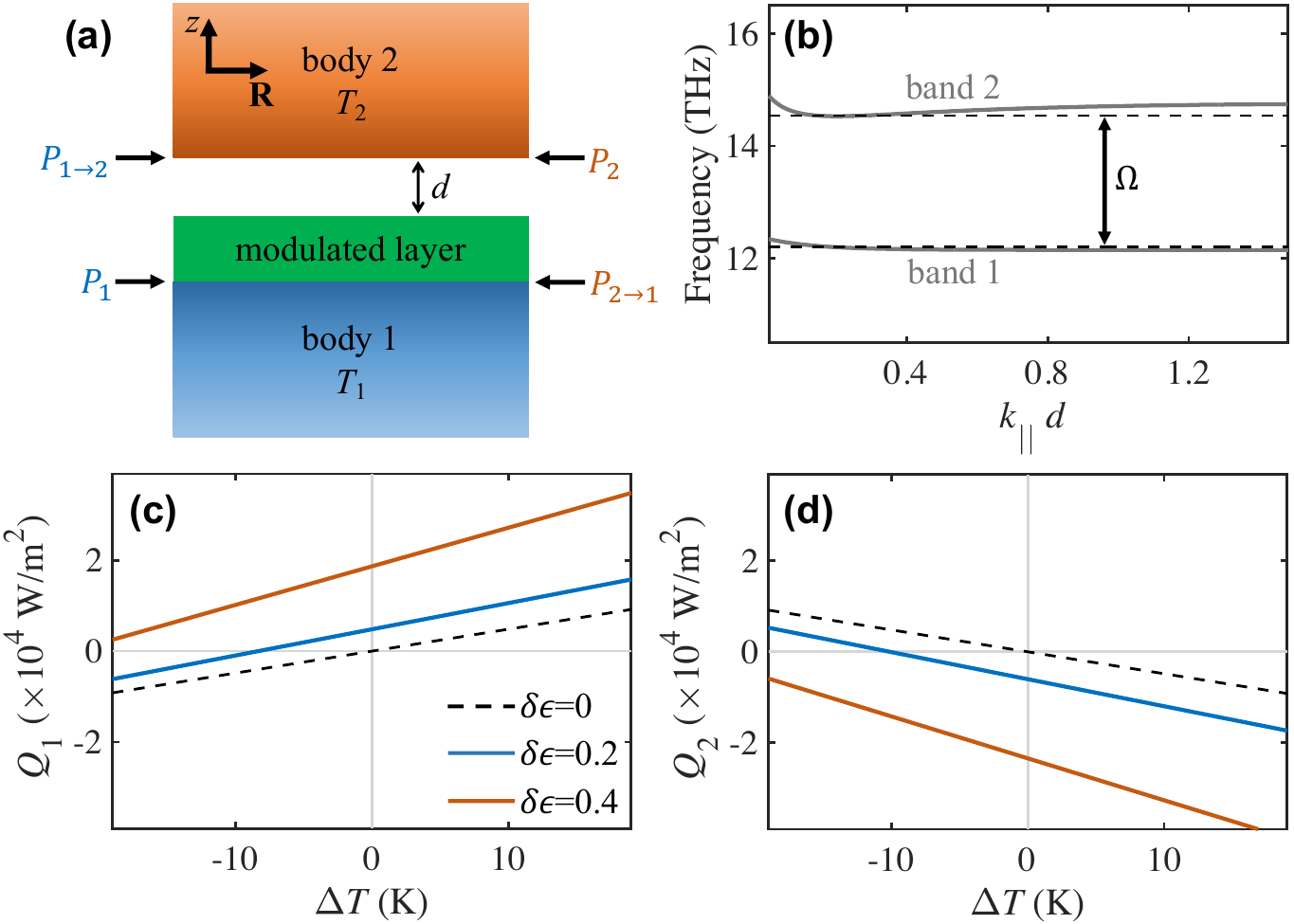}
		\par\end{centering}
	\caption{(a) Schematic of a photonic system composed of two semi-infinitely extended planar structures separated by a vacuum gap of thickness $d=1\,$nm. In the bottom structure, a time-modulated layer (green region) is on top of a substrate (body 1 maintained at temperature $T_1$, blue region), whereas the upper structure consists of only a substrate (body 2 at temperature $T_2$, red region). Black arrows indicate horizontal planes where the Poynting fluxes are calculated. (b) Photonic band structures of the designed system shown in panel (a), where two quasi-flat bands can be seen. The modulation frequency $\Omega$ is set to be the frequency difference between the two bands. (c) Net heat flux $Q_1$ emitted from body 1 as a function of the temperature variation $\Delta T$ between the two bodies, for different modulation strengths $\delta\epsilon$. (d) Same as panel (c) but for net heat flux $Q_2$ emitted from body 2. }
	\label{Fig1}
\end{figure}

As an exemplary system, we consider two semi-infinitely extended planar structures separated by a vacuum gap of thickness $d=1\,$nm in Fig.\ \ref{Fig1}(a). The entire system is translationally invariant along the in-plane direction ${\bf R}\equiv(x,y)$. The bottom structure consists of a lossless time-modulated layer (green region) on top of a substrate (body 1, blue region), whereas the top structure consists only of a substrate (body 2, red region). Bodies 1 and 2 are made of two different polar materials supporting surface phonon polaritons, the permittivities of which are given by $\epsilon_{1,2}(\omega)=\epsilon_{1,2}^{\infty}\left(1+\frac{(\omega_{1,2}^{L})^2-(\omega_{1,2}^{T})^2}{(\omega_{1,2}^{T})^2-\omega^2-i\gamma_{1,2}\omega}\right)$ with $\omega$ the frequency, $\hbar \omega_{1}^{L}=$55\,meV, $\hbar \omega_{2}^{L}=$64\,meV, $\hbar \omega_{1}^{T}=$49\,meV, $\hbar \omega_{2}^{T}=$58\,meV, $\hbar\gamma_1=\hbar\gamma_2=$0.2\,meV, and $\epsilon_1^{\infty}=\epsilon_2^{\infty}=1$. The two bodies are maintained at temperatures $T_1$ and $T_2$, respectively. The permittivity of the time-modulated layer is $\epsilon_3(t)=\epsilon_s+\delta\epsilon\,{\rm cos}(\Omega t)$, with $\epsilon_s=4$ the static permittivity, $\Omega$ the modulation frequency, $\delta\epsilon$ the modulation strength, and $t$ the time. The thickness of the time-modulated layer is assumed to be 4\,nm. 

The system shown in Fig.\ \ref{Fig1}(a) are designed to support two rather flat bands for a broad range of in-plane wave vectors $\kparb\equiv(k_x,k_y)$, as presented in Fig.\ \ref{Fig1}(b). The band at a lower frequency around $\Omega_1=2\pi\times 12.21\,$THz (band 1) corresponds to the surface phonon polariton supported by body 1, whereas the other band at a slightly higher frequency around $\Omega_2=2\pi\times 14.54\,$THz (band 2) corresponds to the surface phonon polariton supported by body 2. The near-field thermophotonic response of our system is dominant by these two polariton modes. We choose the modulation frequency $\Omega=\Omega_2-\Omega_1=2\pi \times 2.33\,$THz so that an efficient interband photonic transition \cite{YF09} can occur for a broad range of $\kparb$. 

The heat flux in the system arises from the radiation emitted by the fluctuating current sources in both bodies. The heat flux $P_{1\rightarrow2}$ from body 1 to body 2 is sourced from the fluctuating currents residing in body 1 (occupying spatial coordinates $\rb_1$) and absorbed by body 2 (occupying spatial coordinates $\rb_2$). It can be calculated by evaluating Poynting fluxes at the surface of body 2. The heat flux $\Pot$ is generally different from the heat flux $P_1$ emitted from body 1, sourced from its fluctuating currents, which can be calculated by evaluating the Poynting fluxes at the surface of body 1. This is because the time modulation can perform work on the electromagnetic field and hence inject or remove energy from the emitted electromagnetic field. By interchanging $1\leftrightarrow2$, we obtain similar definitions for $P_{2\rightarrow1}$ and $P_2$. The net heat flux $Q_1$ emitted by body 1 can be calculated as $Q_1=P_1-P_{2\rightarrow1}$. 

We develop a fluctuational electrodynamics formalism \cite{YF23} to account for two-body radiative heat transfer under time modulation. The net heat flux $Q_1$ is
\begin{align}
	Q_1=Q'_1-E_1, \label{eq:Q1}
\end{align}
with
\begin{align}
	Q'_1&= \sum_l \int_0^{+\infty} d\omega \hbar\omega  \left[n_1(\omega)-n_2(\omega_l)\right] \Fto(\omega,\omega_l), \label{eq:Qp} \\
E_1&=\sum_l \int_0^{+\infty} d\omega \hbar \omega \left[n_1(\omega_l)-n_1(\omega)\right] \Foo(\omega,\omega_l), \label{eq:E1}
\end{align}
where $n_{1,2}(\omega)=\left[{\rm exp}(\hbar \omega/k_{\rm B}T_{1,2})-1 \right]^{-1}$ is the Bose-Einstein distribution function with $\hbar$ the reduced Planck constant and $k_{\rm B}$ the Boltzmann constant, $\omega_l=\omega+l\Omega$ (with $l$ an integer) is the converted frequency, and
\begin{align}
	F_{\alpha\rightarrow \beta}(\omega,\omega_l)=\frac{2\epsilon_0^2}{\pi}\frac{1}{A}\epsilon''_{\beta}(\omega)   \epsilon''_{\alpha}(\omega_l) \nonumber \\
	\times \sum_{i,j}\int_{V_{\beta}} d\rb_{\beta} \int_{V_{\alpha}} d\rb_{\alpha} \left| G'_{ij}(\rb_{\beta},\rb_{\alpha};\omega,\omega_l) \right|^2 \label{eq:F12}
\end{align}
with $\epsilon_0$ the vacuum permittivity, $\epsilon''_{\alpha,\beta}(\omega)={\rm Im}\left \{ \epsilon_{\alpha,\beta}(\omega) \right\} $ ($\alpha,\beta=1,2$ for body 1 or 2), $V_{\alpha,\beta}$ the volume occupied by body $\alpha$ or $\beta$, and $A$ the surface area of the entire structure in the $x-y$ plane. In Eq.\ \ref{eq:F12}, $G'_{ij}(\rb_{\beta},\rb_{\alpha};\omega,\omega_l)$, with $i,j=x,y,z$, is the element of the Green's function that relates the thermally emitted electric fields to the polarization density sources in time-modulated systems. We further define $F_{\alpha\rightarrow \beta}^{(l)}(\omega) \equiv F_{\alpha\rightarrow \beta}(\omega,\omega_l)$. Here, the non-negative $F_{\alpha\rightarrow \beta}^{(l)}$ can be regarded as the photon number flux spectrum associated with the frequency conversion process from $\omega_l$ to $\omega$ when heat is radiatively transferred from body $\alpha$ to $\beta$. As shown in Eq.\ \ref{eq:F12}, multiple up- and down-conversion processes can occur during the radiative heat transfer, corresponding to the $l<0$ and $l>0$ components, respectively. In contrast, only $l=0$ component is non-zero in static systems. Due to the appearance of these energy conversion processes, $Q_1$ in time-modulated systems can be expected to be different from that in static systems. The term $E_1$ depends on the temperature $T_1$ of body 1, and vanishes in static systems. By interchanging $1\leftrightarrow2$ in Eqs.\ \ref{eq:Q1}--\ref{eq:E1}, the quantities $Q_2$, $Q_2'$, and $E_2$ can be obtained. 

\begin{figure*}
	\noindent \begin{centering}
		\includegraphics[width=0.96\textwidth]{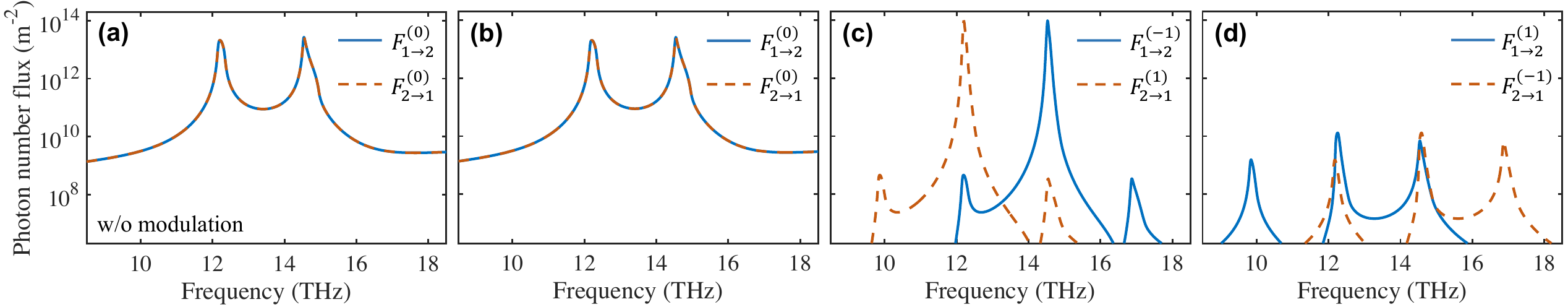}
		\par\end{centering}
	\caption{(a) Photon number flux spectral densities $\Fot^{(0)}$ and $\Fto^{(0)}$, defined in Eq.\ \ref{eq:F12}, for the unmodulated system. (b-d) Time-modulated photon number flux spectral densities $\Fot^{(l)}$ and $\Fto^{(l)}$, for $l=0$ (b) and $l=\pm 1$ (c,d), with $\delta\epsilon=0.4$.}
	\label{Fig2}
\end{figure*}

In Fig.\ \ref{Fig1}(c), we show $Q_1$ as a function of the temperature variation $\Delta T=T_1-T_2$, where $T_1=T_0+\Delta T/2$ and $T_2=T_0-\Delta T/2$ with $T_0=300\,$K, for different modulation strengths $\delta\epsilon$. Without time modulation ($\delta \epsilon=0$, black dashed curve), $Q_1$ is positive (negative) when $\Delta T>0$ ($\Delta T<0$), and vanishes at equilibrium. With time modulation, the behavior of $Q_1$ is drastically altered as an external work $W$, done through the modulation, is applied to the system. When $\delta\epsilon=0.2$ (blue solid curve), we find $Q_1$ is shifted upwards, and it is non-zero when $\Delta T=0$. This upwards shift indicates up-conversion processes are dominant in the heat transfer from body 1 to 2 as we further elaborate below. In the regime where $\Delta T>0$, an enhancement of $Q_1$ is seen as compared with the static case. This enhancement is supplied by the external work. In the regime where $-8.3\,{\rm K} \apprle\Delta T<0$, $Q_1$ remains positive, which means that body 1 experiences active cooling. It has a positive net outgoing heat flux in spite of the fact that it has a temperature lower than that of body 2. In the regime where $\Delta T\apprle -8.3\,$K, $Q_1$ becomes negative, and its magnitude is smaller than that of the static case. Therefore, in this regime time modulation suppresses the heat flux entering body 1.  When $\Delta T\approx-8.3\,$K, $Q_1 = 0$, which means a complete thermal isolation of body 1 from body 2, in spite of the fact that they are in close proximity with each other and have different temperatures. When increasing $\delta\epsilon$ to 0.4 (red solid curve), $Q_1$ is shifted further upwards. This upward shift results in a larger enhancement of $Q_1$ when $\Delta T>0$, as well as a wide range of $\Delta T$ in which active cooling of body 1 occurs.

For the same system, the behavior of net heat flux $Q_2$ out of body 2 is shown in Fig.\ \ref{Fig1}(d).  Without modulation, we have $Q_2 (\Delta T) = - Q_1 (\Delta T)$, as expected from energy conservation arguments. Under modulation, $Q_2$ is shifted downwards from the unmodulated case. As a result, we can also find thermal isolation, suppression, enhancement, and heat pumping for body 2 within different $\Delta T$ ranges. Under time modulation, energy conservation also holds as $Q_1+W=-Q_2$ for each $\Delta T$, with $W=P_{1\rightarrow2}-P_1+P_{2\rightarrow1}-P_2$.

Below, we discuss the physical mechanism underlying the results shown in Fig.\ \ref{Fig1}. Typically, $E_1$ ($E_2$) is much smaller than $Q'_1$ ($Q'_2$). We therefore focus on $Q'_1$ and $Q'_2$. Here, we examine some of the general symmetry properties of $\Fot^{(l)}$ and $\Fto^{(l)}$, which underly the calculation of $Q'_1$ and $Q'_2$ as shown in Eq.\ \ref{eq:Qp}. We first find that the time-modulated Green's function for our system is symmetric, i.e., 
\begin{align}
	G'_{ij}(\rb_{\beta},\rb_{\alpha};\omega,\omega_l)=G'_{ji}(\rb_{\alpha},\rb_{\beta};\omega_l,\omega). \label{eq:G}
\end{align} 
From Eqs.\ \ref{eq:F12} and \ref{eq:G}, we can obtain
\begin{align}
	\Fot^{(l)}(\omega)=\Fto^{(-l)}(\omega_l). \label{eq:Fsym}
\end{align}
This detailed balance relation states that an up- (or a down-)conversion process, for example, from $\omega_l$ to $\omega$, that occurs during the radiative heat transfer from body 1 to 2 is balanced by a down- (or an up-)conversion process from $\omega$ to $\omega_l$ for the heat transfer in the reversed direction.

We show the spectra of $\Fot^{(0)}$ and $\Fto^{(0)}$ in Fig.\ \ref{Fig2}(a) for the unmodulated system. The spectra exhibit two peaks associated with the two surface phonon polariton bands shown in Fig.\ \ref{Fig1}(b), and $\Fot^{(0)}$ coincides with $\Fto^{(0)}$ as expected for passive systems. The spectra of modulated $\Fot^{(l)}$ and $\Fto^{(l)}$ are presented in Fig.\ \ref{Fig2}(b-d) with $l=0,\,\pm1$ for $\delta\epsilon=0.4$. Note that the components with $|l|>1$ have negligible contributions in the radiative heat transfer. For the modulated system, we find that $\Fot^{(0)}$ also coincides with $\Fto^{(0)}$, as predicted by Eq.\ \ref{eq:Fsym}, as shown in Fig.\ \ref{Fig2}(b). The two-peak feature in Fig.\ \ref{Fig2}(b) is almost identical to that for the unmodulated system but with slightly smaller peak values, and shares the same physical origin. 

In Fig.\ \ref{Fig2}(c), we plot $\Fot^{(-1)}$ and $\Fto^{(1)}$. They are identical in terms of the spectral shape but is shifted in frequency by $\Omega$. We note that the spectra are plotted with respect to the frequency of the absorbed photons at the receiving body. The observation here provides a direct check of Eq.\ \ref{eq:Fsym}.  $\Fot^{(-1)}$ exhibits three peaks with the dominant one in the middle near the frequency $\Omega_2$. The dominant peak value of $\Fot^{(-1)}$ is much larger than that of $\Fot^{(0)}$, indicating the significant contribution of the the up-conversion process from $\Omega_1$ to $\Omega_2$ when heat is transferred from body 1 to 2, as a result of the parallel band structure shown in Fig.\ \ref{Fig1}(b). The high value of this peak can be understood by examining Eq.\ \ref{eq:F12}, which involves the factor $ \epsilon''_1(\Omega_1) \epsilon''_2(\Omega_2)$ that is related to $\epsilon''_{1,2}(\omega)$ evaluated at their respective resonance frequencies. We have $\Fot^{(-1)}(\Omega_2) \gg \Fot^{(0)}(\Omega_2)$ since the latter involves a factor of $\epsilon''_1(\Omega_2)  \epsilon''_2(\Omega_2) $, where $\epsilon''_1(\Omega_2) $ is off-resonance. The same observation also holds for $\Fto^{(1)}$, showing an efficient down-conversion from $\Omega_2$ to $\Omega_1$ (with the dominant peak around $\Omega_1$) when the heat flux flows along the opposite direction. The other two minor peaks in $\Fot^{(-1)}$ are associated with the transitions $(\Omega_1-\Omega) \rightarrow \Omega_1$ and $\Omega_2 \rightarrow (\Omega_2+\Omega)$, respectively, which are rather weak because no polariton mode is available at either $\Omega_1-\Omega$ or $\Omega_2+\Omega$. 

In Fig.\ \ref{Fig2}(d), we plot the spectra of $\Fot^{(1)}$ and $\Fto^{(-1)}$. Similar to the cases in Fig.\ \ref{Fig2}(c), these two spectra are the same except for a frequency shift of $\Omega$. Note that the down-conversion component $\Fot^{(1)}$ is orders-of-magnitude smaller than the up-conversion component $\Fot^{(-1)}$ mainly because the down-conversion process from $\Omega_2$ to $\Omega_1$ (the middle dominant peak in $\Fot^{(1)}$) is associated with the factor of $\epsilon''_1(\Omega_2) \epsilon''_2(\Omega_1)$, which is much smaller than the factor $\epsilon''_1(\Omega_1) \epsilon''_2(\Omega_2)$ appearing in $\Fot^{(-1)}$ for the up-conversion process. Thus, we find that the heat transfer from body 1 to 2 (2 to 1) is dominant by the first-order up-conversion (down-conversion) process, induced by the time modulation, which results in the unusual behavior of the near-field heat transfer as discussed in Figs.\ \ref{Fig1}(c,d).

\begin{figure}
	\noindent \begin{centering}
		\includegraphics[width=0.48\textwidth]{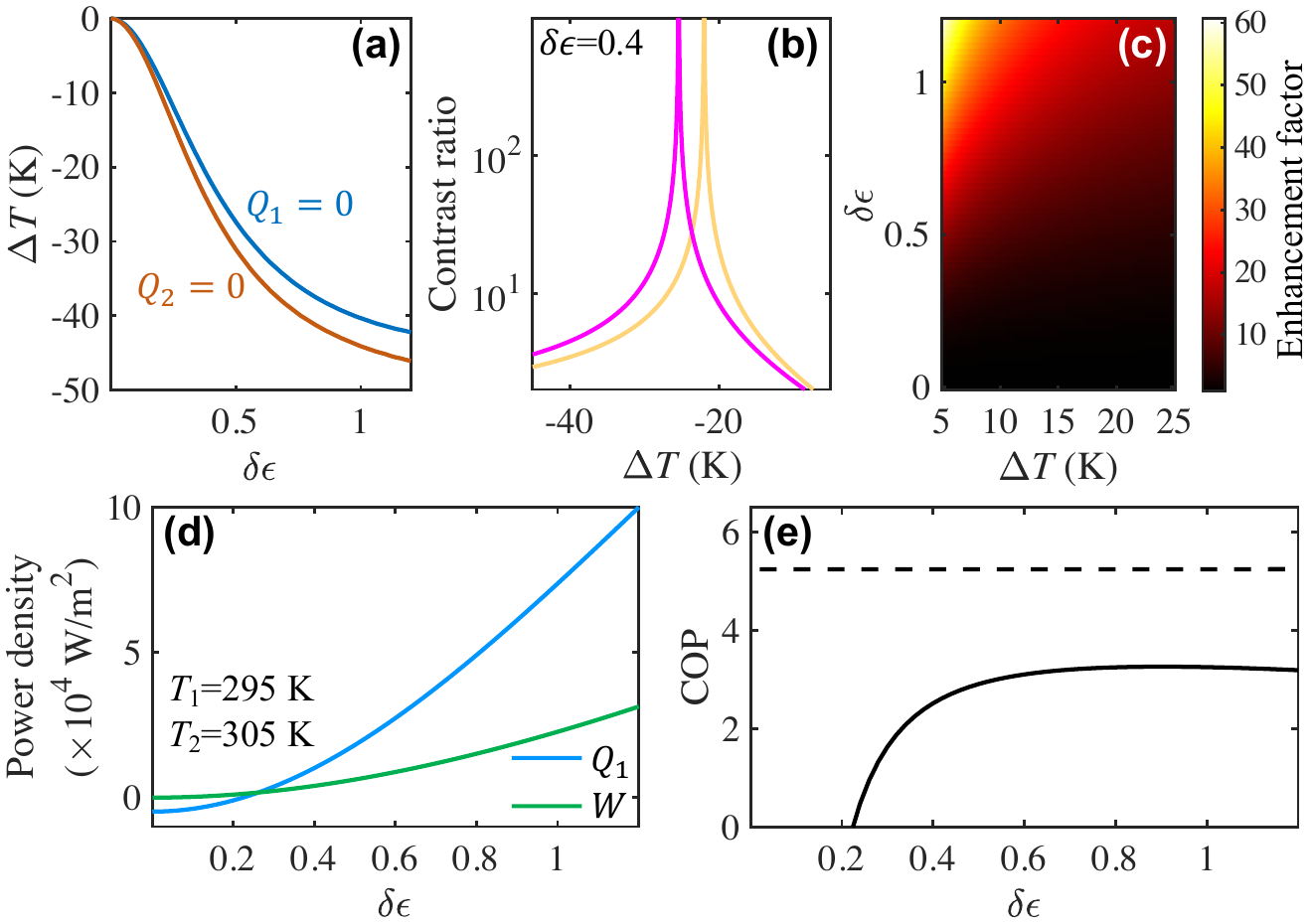}
		\par\end{centering}
	\caption{(a) Temperature variation $\Delta T$ as a function of the modulation strength $\delta \epsilon$ for the system, displayed in Fig.\ \ref{Fig1}, at which $Q_1=0$ (blue curve) or $Q_2=0$ (red curve). (b) Contrast ratio, defined by $|Q_{1,2}(- \Delta T)/Q_{1,2}(\Delta T)|$, as a function of $\Delta T$ at a fixed $\delta\epsilon=0.4$, represented by orange and magenta curves, respectively. (c) Enhancement factor of $Q_1$ as a function of both $\Delta T$ and $\delta \epsilon$ for the system operating in the amplification regime with $\Delta T>0$. (d) Net heat flux $Q_1$ (blue curve) and work power density $W$ (green curve) as a function of $\delta \epsilon$ for the system operating in the active cooling regime (when $Q_1>0$) with $T_1=295\,$K and $T_2=305\,$K. (e) The COP (black solid curve), given as $Q_1/W$, for the radiative refrigeration presented in panel (d). Black dashed curve represents the upper bound of COP.}
	\label{Fig3}
\end{figure}

We now discuss various implications of near-field heat transfer between time-modulated media. A key observation in Figs.\  \ref{Fig1}(c,d) is the possibility of thermal isolation, where the heat transfer into one of the bodies is eliminated even when $\Delta T \neq 0$. For many applications \cite{CSF15,CSS15}, it is important to eliminate the near-field heat transfer as a parasitic pathway. In Fig.\ \ref{Fig3}(a), we plot the temperature difference $\Delta T$, at which the heat transfer into either bodies vanishes, as a function of modulation strength $\delta \epsilon$. For a larger $|\Delta T|$, a stronger modulation is required to eliminate the heat transfer.

In Figs.\ \ref{Fig1}(c,d), in the presence of modulation, the magnitudes of both $Q_1$ and $Q_2$ are asymmetric with respect to $\Delta T$, i.e., $|Q_{\alpha}(\Delta T)| \neq |Q_{\alpha}(-\Delta T)|$ ($\alpha=1$ or 2). Therefore, the structure operates as a thermal rectifier or diode. Previous works on radiative thermal diodes are based on temperature dependent refractive indices of the bodies \cite{OLF10,BF11,WZ13,BB13,JED15,FTZ18,OMB19,KOJ20,ZZY20,LBN21,LHC21}. The use of modulation thus represents a novel mechanism for achieving thermal rectification. 
The performance of a thermal diode can be characterized by the contrast ratio $|Q_{1,2}( -\Delta T)/Q_{1,2}( \Delta T)|$. In Fig.\ \ref{Fig3}(b), we show $|Q_1(-\Delta T)/Q_1(\Delta T)|$ (orange curve) as a function of $\Delta T$ for a fixed modulation strength $\delta\epsilon=0.4$. In particular, this quantity diverges at $\Delta T\approx -22\,$K since $Q_1 (\Delta T)$  = 0, whereas $Q_1 (-\Delta T) \neq 0$ at this temperature bias, which results in a diode of infinite contrast ratio. Similar behavior of infinite contrast ratio can be seen in $|Q_2(-\Delta T)/Q_2(\Delta T)|$ as well. The obtained infinite contrast ratio can not be achieved in previous designs of thermal diodes \cite{OLF10,BF11,WZ13,BB13,JED15,FTZ18,OMB19,KOJ20,ZZY20,LBN21,LHC21}, and points to unique capabilities of time modulation for active control of near-field radiative heat flow.

Figures\ \ref{Fig1}(c,d) show the existence of a temperature range where the heat flux is enhanced by the modulation. Various methods have been previously explored to improve the power density of near-field heat transfer \cite{FMV08,LZZ14,FGF17,VMJ20,KZZ20,LZ14,ZGZ17,ZM20,CSS15}.  In our time-modulated system, we 
define an enhancement factor for $Q_1$ as $|Q_1(\Delta T, \delta \epsilon)/Q_1(\Delta T, \delta \epsilon = 0)|$, and plot it as a function of both $\Delta T$ and $\delta\epsilon$ in Fig.\ \ref{Fig3}(c). For a given $\Delta T$, the enhancement factor increases as $\delta\epsilon$ increases. A maximum of $\sim 60$-fold enhancement is found for the parameter range used here with $\Delta T=5\,$K. The results indicate the significant capability of time modulation to enhance near-field radiative heat flow.

Finally, Figs. \ref{Fig1}(c,d) indicate a regime of active cooling in our system. To explore this regime further, we plot $Q_1$ as a function of the modulation strength $\delta\epsilon$ in Fig.\ \ref{Fig3}(d), for the case where $T_1=295\,$K and $T_2=305\,$K.  Radiative refrigeration ($Q_1>0$) starts when $\delta\epsilon \apprge 0.24$. In accordance with the results shown in Figs.\ \ref{Fig1}(c,d), the cooling power density $Q_1$ increases when increasing $\delta\epsilon$. At $\delta \epsilon  = 0.5$, the cooling power density is $\sim 1.8\times10^4\, \rm{W/m^2}$, which is about five orders of magnitude larger as compared with the far-field case considered in Ref.\ \cite{BLF20}, at a similar modulation strength and temperature difference between the two bodies, thus demonstrating a significantly enhanced cooling power density in the near-field regime. The work $W$ done through the modulation is shown as a function of $\delta\epsilon$ in Fig.\ \ref{Fig3}(d), and it also increases as $\delta\epsilon$ increases. In the active cooling regime, the coefficient of performance (COP) of our system is given by $Q_1/W$, which is plotted as a function of $\delta\epsilon$ in Fig.\ \ref{Fig3}(e) (black solid curve). The upper bound of COP for our system can be proved to be ${\Omega_1}/{\Omega}\approx 5.24$ [black dashed curve in Fig.\ \ref{Fig3}(e)], which is related to the Carnot limit $T_1/(T_2-T_1)$.

As final remarks, in our theoretical model we have chosen a small gap size of 1\,nm, which has been realized in near-field heat transfer experiments \cite{KSF15}, to provide an assessment of the upper range of reachable power densities in our schemes. Moreover, all demonstrated novel effects in our work can be achieved with a larger gap size as well as with a thicker time-modulated layer, where nonlocal effects should not play a significant role \cite{CVH08}. Our theoretical model can be implemented in practice by using indium phosphide \cite{P1985} and quartz \cite{DMD12} for bodies 1 and 2, respectively, and low-loss nonlinear dielectric materials \cite{CLP19,XTM22} can be used for constructing the time-modulated layer.


In summary, we have shown that a time-modulated photonic system can be used to achieve radiative heat flux amplification, active cooling, and thermal isolation. Extending our demonstrated concepts towards the quantum regime, such as quantum thermal machines \cite{LBA15_2}, is also of future interests. Our findings open a promising avenue toward perfect radiative thermal diodes as well as nanoscale thermal energy harvesting and management.

\begin{acknowledgments}
This work has been supported by a MURI program from the U. S. Army Research Office (Grant No. W911NF-19-1-0279).
\end{acknowledgments}


\end{document}